\documentclass[10pt]{article}
\usepackage{graphicx}

\usepackage{epsfig}
\usepackage{latexsym}
\usepackage{amsmath}
\begin{document}

\title{Complex-valued information entropy measure\\ for  networks with directed links (digraphs).\\   Application to   citations  by community agents with opposite  opinions 
 }
\author{ G. Rotundo\footnote{ 
  $e$-$mail$ $address$: giulia.rotundo@uniroma1.it} \\Faculty of Economics, University of Tuscia,\\ via del Pa\-ra\-di\-so 47,  I-01100 Viterbo, Italy.\\ {\it present address: }\\
  Department of Methods and Models for Economics, Territory and Finance\\
Sapienza University of Rome\\
via del Castro Laurenziano 9
00161 Rome, Italy
 \\
  and \\
 M. Ausloos\footnote{$e$-$mail$ $address$:
marcel.ausloos@ulg.ac.be} \\   R\' esidence Beauvallon,
 rue de la Belle Jardini\`ere,  483/0021 \\B-4031, Angleur, Belgium\\
 {\it  previously at } GRAPES@SUPRATECS, Universit\'e
de Li\`ege,  \\Sart-Tilman, B-4000 Li\`ege, Euroland.
   }

\date{\today}
\maketitle
\vskip 0.5truecm

\begin{abstract}
The notion of  complex-valued  information entropy measure  is presented. It applies in particular  to directed networks (digraphs).  
The corresponding statistical physics notions are outlined.  The  studied network, serving as a    case study,  in view of  illustrating  the discussion, concerns citations by agents belonging to two distinct communities which have markedly different opinions:   the  Neocreationist and Intelligent Design  Proponents,  on one hand, and   the Darwinian  Evolution Defenders,  on the other hand.  \newline  The     whole,   intra- and inter-community  adjacency matrices, resulting from  quotations of published work by the community agents,  are elaborated and eigenvalues calculated.  Since eigenvalues can be complex numbers, the information   entropy may become also complex-valued. It is calculated for the illustrating case. \newline The role of the   imaginary part finiteness  is discussed in particular and given some  physical sense interpretation through local interaction range consideration. It is concluded  that such generalizations  are not only interesting and necessary for discussing directed networks, but also may  give new  insight into conceptual ideas about directed or other networks. Notes on extending  the above to Tsallis entropy measure are found in an Appendix.

\end{abstract}
keywords : Entropy -- Citation networks -- Asymmetric adjacency matrix


\maketitle

\section{
Introduction } \label{intro}
 
     "Complicated systems" are usually, but abusively \cite{Procaccia},  called complex systems.  
 Recall that complex numbers are found  in physics to describe various macroscopic properties:   the  dielectric permittivity,   the  electric  impedance, the amplitudes and phase angles of  modal vibrations, Magnetic Resonance images, etc..
  
  Complex eigenvalues (EVs) do  $naturally$ occur  in a Hamiltonian formalism:  the imaginary part of some self-energy  which turns out to be the "density of states", localization in superconductors \cite{r6}, dissipation and scattering in Quantum Chaos \cite{r7a,r7b} or Quantum Chromodynamics,  with a non-vanishing chemical potential \cite{r[8]},  fractional Quantum Hall Effect  \cite{r[9]},  two-dimensional plasma of charged particles \cite{r10,r11}.   See also the imaginary part of  the  "free energy"   measuring the quantum decay rate of a pure  state or the imaginary part of  power law exponents  describing oscillations in the specific heat or  in the electrical resistivity temperature derivative  at (magnetic, for example) transitions  \cite{MAapl43,MAriste}.
  
  It will be argued below that  one modern case of interest is the entropy of {\it directed networks}: it  can have a real and an imaginary part.   
   For coherence, the definitions of  the  thermodynamic 	and the information entropy measure  are recalled   in Sect.  \ref{CIE}.
Subsequently   the  complex information entropy measure (CIE) is defined in order to take into account complex algebra, when or if necessary, - in particular when discrete "states"  are considered.     In Sect. \ref{exempldata},  a brief outline of some  illustrating case is presented. It concerns a citation network,   - see Sect. \ref{2comm}.  The entropy of such  digraphs   can be complex, following some illustrative calculation. A preliminary interpretation follows in Sect. 
    \ref{discussion}.

  Some conclusion is found in Sect. \ref{concl}. A discussion on normalization is found in Appendix A. An extension to Tsallis entropy measure is  suggested in  Appendix B. 
 A long but useful discussion on the origin of complex EV for citation and similar asymmetric networks is found in Appendix C, where the various  cases of the most simple asymmetric networks, i.e. triads,  described by 3x3 matrices, are used to point to  transitivity causes.

\section{Complex-valued information entropy measure}\label{CIE}

The typical statistical mechanics approach starts from the partition function $Z$ defined  through the sum of all the Boltzmann factors measuring the probability of occurrence of the various states of a Hamiltonian $\cal H$. 
The partition function   reads
 \begin{equation} \label{Z}
  Z_s(T) =  \Sigma_{\nu=1}^{N_s}  e^{- \beta \cal H(s, \nu) }.
 \end{equation}
when observing a system at a scale $s$ and at some temperature $T$ ($\beta\equiv 1/k_BT$); $\nu$  $(= 1, ..., N_s)$ is an index allowing to label and to count the configurations of the relevant degrees of freedom governing the system. One can next derive the corresponding free energy $F_s(T)= -k_BT \;ln (Z_s(T))  $. This free energy is known to be a homogeneous function at critical points   \cite{stanley},  
   in the scaleless situation. This leads to  define (real) critical exponents \cite{FisherRMP,StanleyRMP}.  The scale of interest is in fact related to the coherence length, $\xi$,  itself  temperature dependent, giving some measure of the "temperature distance", $\epsilon \equiv (T-T_c)/T_c$, between the critical temperature $T_c$ and the system temperature $T$.
Practically,  one  can map  the  model Hamiltonian $\cal H$  over the possible states, through $e^{ -\beta \cal H}$.  Without justifying much here, i.e. going back to the theory of Markov processes, we admit that the latter is
 a so called transfer matrix \cite{Thompson}, with elements made of Boltzmann factors. Thereafter we can  calculate the eigenvalues $\lambda_{i}$ and   evaluate  $Z$, $F$, and $S$, i.e.
$e^{ -\beta {\cal H}}$  $ \rightarrow  $ $e^{ -\beta E}$ $ \rightarrow  $ $e^{ -\beta \lambda_{i}}$
$\rightarrow  $ $Z$ 
$ \rightarrow$ $F$     
$ \rightarrow$ $\simeq$ $\lambda_{1}-k_BT\; ln\left( 1+ e^{-\beta( \lambda_{2}-  \lambda_{1})}   + ...\right)$            
 $ \rightarrow$ $S$.
 Often, the first eigenvalue $\lambda_{1}$ is sufficient to obtain an estimate of the free energy.  So called "corrections" imply higher order terms \cite{brezin}, though one often stops at  $\lambda_{2}$. 
Thus,  any (thermodynamic) property,  $\chi(s, T) $,  derived from $F_s(T) \sim  F (\xi(\epsilon )) $, can have the form 
$\chi(\epsilon) \sim  \xi^{\tau}(\epsilon), $
 but  where $\tau$  can be complex \cite{DSI,DSphysrep}.   In some sense, this is as if $F$ is complex, or if some  $\lambda_{i}$  is complex.  

We interpret an adjacency matrix, e.g. describing a  network, as a transfer matrix, - written in terms of some Hamiltonian $\cal H$.  However, we are aware that such an $\cal H$ is not necessarily a {\it bona fide} Hamiltonian,  i.e., from a  quantum mechanics point of view,  since it might not be hermitian. The eigenvalues can thus be complex.
 
\subsection{Real algebra information entropy measure}\label{RIEM}

 The  Boltzmann entropy   in statistical physics ~\cite{Huang} reads:  
\begin{equation} S=- \int_0^{\infty} \mbox{d}w N(w,t)\log{N(w,t)}. \end{equation} 
where $ N(w,t)$  is the number of states, of type $w$, at time $t$.  This entropy   corresponds to the usual Shannon information entropy which can also be written in terms of  the (equilibrium) probability  $p_i$   of  finding some (macroscopic variable in some) state $i$,
 \begin{equation} \label{shannon} H= - k_B  \sum_i p_i \ln p_i
\end{equation}  Since $p_i$ is necessarily positive but  less than 1, $H \ge 0$, - like a Boltzmann factor.  

Often, $p_i$  is  practically known through some {\it a priori} bin decomposition
of the distribution of states. This is similar to the Theil approach in economy \cite{Theil,Theil65,Misk2008a}, where the individual incomes are taken into account, normalized to the income of the whole population of interest in the study.  Thus, in  the above,  some generalization would replace $p_i$ by $\lambda_i$. 

Having in mind applications to network theory, one might argue that the most adequate normalization should take into account the size of the system (or the adjacency matrix). In the Theil spirit, one could imagine various normalizations, see Appendix A.  This can be considered elsewhere depending on specific applications.

 Thereafter, the  \textit{Information  Entropy} ($IE$)  is defined, 	as   in \cite{medy,wang}, by 
\begin{equation}
\label{in}
H= 1 + \sum\limits_{i = 1}^{l_M} {\frac{{\lambda _i }}{{l_M}}\log _{l_M} \frac{{\lambda _i }}{{l_M}}}
\end{equation}
in units where $k_B=1$, and where $\lambda_i$ is the \textit{i-th} eigenvalue of the  relevant   matrix of size $l_M$ x $l_M$.  However, one may, without much loss of   generality, decide that the log-basis is the natural one, rather than one being associated with the matrix size, i.e. to keep a thermodynamic-like spirit,  and redefine for our purpose, 

\begin{equation}
\label{in2}
H_e \equiv 1 + \sum\limits_{i = 1}^{l_M} {\frac{{\lambda _i }}{{l_M}}\ln \frac{{\lambda _i }}{{l_M}}}.
\end{equation}
to emphasize that we the  natural log. Nevertheless, for simplicity, the index $_e$ will not be further written.

In concluding this section, let us introduce the  notation 
$H_1 \equiv 1 +   {\frac{{\lambda _1}}{{l_M}}\ln \frac{{\lambda _1}}{{l_M}}}.$
such that it should correspond to an estimate of the  $IE$ based on the largest EV.

\subsection{Complex information entropy measure}\label{CIEM}

In fact, let $l_M$ is the number of available states, or the number of EVs for some Hamiltonian or transfer matrix  or adjacency matrix. Since the $l_M$ eigenvalues could be complex, one has to generalize the previous formula, Eq. (\ref{in2}) to the complex plane, i.e. $H=  H^{'}+\;i\;H^{''}$,  such as one gets

 \begin{equation}
\label{rin2}
 H^{'}= 1 + \sum\limits_{i = 1}^{l_M}\;  \frac{{\|\lambda _i }\|} {{l_M}} \;   \left( [\ln \frac{{\|\lambda _i }\|}{{l_M}}][  cos(\Phi_i)]  -[(\Phi_i+i \;n \;2\pi)\; sin(\Phi_i)]\right),
\end{equation}
and
\begin{equation}
\label{iin2}
H^{''} =  \sum\limits_{i = 1}^{l_M}\;  \frac{{\|\lambda _i }\|}{{l_M}}  \;   \left([ \ln \frac{{\|\lambda _i }\|}{{l_M}}  ][sin(\Phi_i)] +[ (\Phi_i+i \;n \;2\pi)\; cos(\Phi_i)] \right),
\end{equation}
where we identify any  $\lambda _i  ^{'}+i\; \lambda _i   ^{''} $ with  $\| \lambda _i \|\;[ cos (\Phi_i) +i \;sin (\Phi_i)]$.  Note that from these, the magnitude $\|H\|$ of the complex information entropy, as well as the $IE$ phase factor,  could be calculated.   The $i \;n\;2\pi$ term is due to the fact that the argument of  log 
$\lambda _i $   is  a multivalued function  
 \cite{complexalgebra}.  Therefore, for meaningfully pursuing any calculation, it is necessary to define the existence interval  of  the argument of  such a complex number. 
 One can decide that $\Phi_i$ is defined 
 in  $ 0 \le \; \Phi_i\; <  +2\pi$, or   in $ -\pi < \; \Phi_i\; \le \; +\pi$, i.e.  the so called  "principal value" (PV) component   \cite{complexalgebra}.

Thus, when doing the summations in Eq.(\ref{rin2}) and Eq.(\ref{iin2}), several terms may cancel each other.  In particular, when either   $\Phi_i \; tan(\Phi_i)$ is even or when $\Phi_i \; cotan(\Phi_i)$ is odd.  This depends on  whether   two EVs  are complex conjugates, and in the appropriately defined space of  $  \Phi_i$.
Then, the above formulae can be simplified through some trivial algebra when taking into account the form of  the eigenvalue.

  Whence, starting from Eq.(\ref{in2}), let us distinguish between the   real  positive ($\rho$),   real  negative ($\nu$),   imaginary ($  \mu$),  
 equal to zero, and complex-valued ($\lambda)$ eigenvalues.   One can rewrite Eq.(\ref{iin2}), with obvious notations for each sum upper limit,
\begin{equation}
\label{inrhomulambda}
H= 1 + 
\sum\limits_{i = 1}^{\rho_M} \frac{\rho _i }{l_M}     \ln \frac{\rho _i }{l_M}
+\sum\limits_{i = 1}^{\nu_M}  \frac{\nu _i }{l_M}    \ln \frac{\nu _i }{l_M}
+\sum\limits_{i = 1}^{\mu_M}  \frac{ \mu _i }{l_M}      \ln \frac{\mu _i }{l_M}
+\sum\limits_{i = 1}^{\lambda_M} \frac{\lambda _i }{l_M}   \ln \frac{\lambda _i }{l_M}\; .
\end{equation}

The  first sum is trivially real, for any interval definition of $\Phi_i$.  In fact, $ \rho _i  = \|\rho _i \| $.

When an EV has some finite, in particular if it is negative,   real or imaginary part,  several cases   must be distinguished. The case of degenerate EVs has also to be specially considered. Note that if an EV is  evenly degenerate, it can be considered as stemming from a set of $c.c.$ EV with a zero imaginary part. Then,  $\frac{\lambda _i }{l_M}   \ln \frac{\lambda _i }{l_M}$ depends on which complex sheet ("interval space") the phase factor   is defined.


 \subsubsection{   $\Phi_i$ $\in$  [0, 2$\pi$[ space}\label{02pi}
 \begin{itemize}
 \item When  
 $ \nu _i ^{'}  $  = -$\| \nu _i ^{'}\|$, and   $ \nu _i ^{''}=0$,   one has $ \nu _i  \equiv    \| \nu _i ^{'}\|\; e^{i\pi+ 2ni\pi}$; the second summation has thus terms like 
 $ {\frac{{\nu _i }}{{l_M}}\ln \frac{{\nu _i }}{{l_M}}}$ 
 $\equiv$                
$ - \frac{\| \nu _i ^{'}\|}{l_M}  \; \left[ln  [\frac{\| \nu _i ^{'}\|}{{l_M}}] +   (2n+1) i\; [ \pi ]\right]$. Such a term contains a real $and$ an imaginary part. The $n=0$ case is of course in order here below, i.e.   $ {\frac{{\nu _i }}{{l_M}}\ln \frac{{\nu _i }}{{l_M}}}$ 
 $\equiv$                
$ - \frac{\| \nu _i ^{'}\|}{l_M}   \;ln  [\frac{\| \nu _i ^{'}\|}{{l_M}}]  - \frac{\| \nu _i ^{'}\|}{l_M}  i\; [ \pi ]$.
 \item In the third summation,  
 the imaginary part   $ \mu _i ^{''}  $ could be positive or negative, i.e. $\mu _i \equiv i\; \mu _i ^{''}=\pm i \|  \mu _i \|$;  the phase factor is  $\left[ \frac{\pi}{2}\right]  $ and $\left[ \frac{3\pi}{2}\right] $, 
  respectively. 
 Thus,  a term in the third summation reads either
 
$  + i  \frac{\| \mu _i \|}{l_M}\; \left[ ln (\frac{ \| \mu _i \| }{l_M}) + i[ \frac{\pi}{2}]\right]  $$\equiv -[\frac{\pi}{2}]  \frac{\| \mu _i \|}{l_M}\;  +  i   \frac{\| \mu _i \|}{l_M}\;  ln (\frac{ \| \mu _i \| }{l_M})$, 

or
$  - i  \frac{\| \mu _i \|}{l_M}\; \left[ ln(\frac{ \| \mu _i \| }{l_M}) + i[ \frac{3\pi}{2}]\right]  $$\equiv [  \frac{3\pi}{2}]  \frac{\| \mu _i \|}{l_M}\;    $  
$-   i \frac{\| \mu _i \|}{l_M}\;  ln (\frac{ \| \mu _i \| }{l_M})$. 

Therefore, in the summation, extending over the whole number ($\mu_M$) of    imaginary, - but necessarily  ($c.c.$), EVs,  we insist, {\bf only}  a $real$  term subsists in $H$   after  summing over  $c.c.$ EVs, i.e. $+  \| \mu _i \|\; \pi/l_M$. Note the {\bf +}  sign.

\item   Let  $\lambda _i  ^{'} +i\; \lambda _i   ^{''} \equiv \| \lambda _i \|\;[ cos (\Phi_i)  + i \;sin (\Phi_i)] \equiv  \| \lambda _i \|\;  e^{+ i\Phi_i}$. Thus, $\lambda _i  ^{'} -i\; \lambda _i   ^{''} \equiv \| \lambda _i \|\;[ cos (\Phi_i)  - i \;sin (\Phi_i)] \equiv  \| \lambda _i \|\;  e^{+ i(2\pi-\Phi_i )}$.
 One has,

$   \frac{\| \lambda _i ^{}\|}{l_M}\; ln\frac{\| \lambda _i ^{}\|}{l_M} \equiv\; \frac{\| \lambda _i ^{}\|}{l_M}\left [cos(\Phi_i)\; ln\frac{\| \lambda _i ^{}\|}{l_M} -\Phi_i\;sin(\Phi_i) \right]   $\\ \;\;$  + \; \;  i\; \; \frac{\| \lambda _i ^{}\|}{l_M}\;   \left [ \Phi_i\;cos(\Phi_i) +   ln\frac{\| \lambda _i ^{}\|}{l_M} sin(\Phi_i)\right],  $...
   if  $   \lambda\equiv\lambda _i  ^{'}  +i\; \lambda _i   ^{''}$. 
   
   But when   $\lambda\equiv\lambda _i  ^{'} -i\; \lambda _i   ^{''}$, one has
$   \frac{\| \lambda _i ^{}\|}{l_M}\; ln\frac{\| \lambda _i ^{}\|}{l_M} $\\$  \equiv\;  \frac{\| \lambda _i ^{}\|}{l_M}\; . \left[  cos(\Phi_i)\; ln\frac{\| \lambda _i ^{}\|}{l_M} +(2\pi-\Phi_i)\;sin(\Phi_i) \right]$  
   
 +$ i$ $ \frac{\| \lambda _i \|}{l_M}     \left[(2\pi- \Phi_i)\;cos(\Phi_i) - ln\frac{\| \lambda _i ^{}\|}{l_M} \; sin(\Phi_i) \right].$
Therefore, the fourth summation can be simplified;   after grouping $c.c.$ terms, as indicated by the notation  $ \sum^{'}  _{ i} $, it reads
 
$ \sum^{'}  _{ i} $ $2 \frac{ \| \lambda _i \|}{l_M}  \left[ cos (\Phi_i)\; ln  \frac{ \| \lambda _i \|}{l_M} +(\pi- \Phi_i ) sin (\Phi_i)\right]+$ \\  $+  i$ $ \sum^{'} _{ i} \; 2\pi    \frac{\| \lambda _i ^{}\|}{l_M}  cos (\Phi_i) $.
\end{itemize}

  One can rewrite the real and imaginary parts  of the $ IE$ such that,  
  \begin{eqnarray} 
\label{rin2fTC} 
 \phantom{}H^{'}_{TC}=  1 + \sum\limits_{i = 1}^{\rho_M}\;  \frac{{\|\rho _i }\|} {{l_M}} \;   \left[  \ln (\frac{{\|\rho _i }\|}{{l_M}})\right]
 +
\nonumber\\
  \sum\limits_{i = 1}^{\nu_M}\;   \frac{-\|\nu _i \|} {l_M} \;   \left[  \ln (\frac{{\|\nu _i }\|}{{l_M}})\right]
+    \nonumber\\
  \sum _{ i}   ^{'} \;      \frac{ \|\mu _i \|} {l_M} \;   \left[  \pi\right]
+  \nonumber\\
 2 \;    \sum _{ i}  ^{'}     \left[ \frac{ \lambda _i  ^{'} }{l_M}          \ln (\frac{{\|\lambda _i }\|}{{l_M}})  + (\pi- \Phi_i) \frac{\lambda _i   ^{''} } {l_M}\right]  
\end{eqnarray}
and
 \begin{eqnarray} 
\label{iin2fTC} 
H^{''}_{TC} =   \sum\limits_{i = 1}^{\nu_M}\; \left[ \frac{-\|\nu _i \|\;\pi} {l_M} \;    \right] + \sum ^{'} _{ i}  \;2\pi    \frac{\| \lambda _i ^{}\|}{l_M}cos (\Phi_i) 
.\end{eqnarray}

\subsubsection{Principal Value (PV) space  :  $\Phi_i$ $\in$  ]-$\pi, \pi$]}
\label{PV}

\begin{itemize}
\item When  
 $ \nu _i ^{'}  $  = -$\| \nu _i ^{'}\|$, and   $ \nu _i ^{''}=0$,   one can obtain, as  {\bf  in TC space},    $ {\frac{{\nu _i }}{{l_M}}\ln \frac{{\nu _i }}{{l_M}}}$ 
 $\equiv$                
$ - \frac{\| \nu _i ^{'}\|}{l_M}   \;ln  [\frac{\| \nu _i ^{'}\|}{{l_M}}]  - \frac{\| \nu _i ^{'}\|}{l_M}  i\; [ \pi ]$.\item Recall that in the third summation,  
 the imaginary part   $ \mu _i ^{''}  $ could be positive or negative, i.e. $\mu _i \equiv i\; \mu _i ^{''}=\pm i \|  \mu _i \|$. In PV space, the phase factor is   $\left[ \frac{\pi}{2}\right]  $ and $\left[ -\frac{\pi}{2}\right]  $, 
   respectively. Thus,  a term in the third summation reads either
$  + i  \frac{\| \mu _i \|}{l_M}\;  \left[ ln (\frac{ \| \mu _i \| }{l_M}) + i[ \frac{\pi}{2}]\right]  $ 
or
$  - i  \frac{\| \mu _i \|}{l_M}\; \left[ ln(\frac{ \| \mu _i \| }{l_M})   - i[ \frac{\pi}{2}]\right]  $.

Therefore,
in the summation, extending over the whole number ($\mu_M$) of   imaginary, - but necessarily  $c.c.$ EVs,  we again insist, {\bf only}   a $real$  term subsists in   $H$   after  summing over  $c.c.$ EVs, i.e. $-  \| \mu _i \|\; \pi/l_M$.  Note the {\bf $ -$} sign.

\item
Because $\lambda _i  ^{'}\pm i\; \lambda _i   ^{''} \equiv \| \lambda _i \|\;[ cos (\Phi_i)  \pm i \;sin (\Phi_i)] \equiv  \| \lambda _i \|\;  e^{\pm i\Phi_i}$,      terms like  $   \frac{\| \lambda _i ^{}\|}{l_M}\; ln\frac{\| \lambda _i ^{}\|}{l_M} \equiv\;$\\$ \frac{\| \lambda _i ^{}\|}{l_M}\left[ [cos(\Phi_i) \; ln\frac{\| \lambda _i ^{}\|}{l_M} -\Phi_i\;sin(\Phi_i) ] \pm\;i [ \Phi_icos(\Phi_i)   + sin(\Phi_i)\; ln\frac{\| \lambda _i  \|} {l_M}  ] \right]$.

Taking into  account that the   $\lambda$  EVs are necessarily  $c.c.$, the fourth summation can be simplified into a {\bf  real} quantity, after grouping $c.c.$ terms, as indicated by $ \sum^{'}  _{ i} $, i.e.

$ \sum _{i  } ^{'}$
$ 2\left[\frac{\lambda _i  ^{'}} {l_M}\; ln \frac{ \| \lambda _i \|}{l_M} - \Phi_i \;\frac{ \lambda _i  ^{''}}{l_M}\right]$   which also reads

 $ \sum^{'}  _{ i} $ $2 \frac{ \| \lambda _i \|}{l_M}  \left[ cos (\Phi_i)\; ln  \frac{ \| \lambda _i \|}{l_M}  -  \Phi_i sin (\Phi_i)\right]$.

\end{itemize} One can regroup the real and imaginary parts such that   finally, 
 
  \begin{eqnarray} 
\label{rin2fPV} 
 \phantom{}H^{'}_{PV}=  1 + \sum\limits_{i = 1}^{\rho_M}\;  \frac{{\|\rho _i }\|} {{l_M}} \;   \left[  \ln (\frac{{\|\rho _i }\|}{{l_M}})\right]
 +
\nonumber\\
  \sum\limits_{i = 1}^{\nu_M}\;  \frac{-\|\nu _i \|} {l_M} \;   \left[  \ln (\frac{{\|\nu _i }\|}{{l_M}})\right]
+    \nonumber\\
  \sum  _{ i}^{'} \; - \frac{ \|\mu _i \|} {l_M} \;   \left[  \pi\right]
+  \nonumber\\
 2 \;    \sum^{'}  _{ i} \left[ \frac{ \lambda _i  ^{'} }{l_M}          \ln (\frac{{\|\lambda _i }\|}{{l_M}})  -  \Phi_i \frac{\lambda _i   ^{''} } {l_M}\right]  
\end{eqnarray}
and
 \begin{eqnarray} 
\label{iin2fPV} 
H^{''}_{PV} =   \sum\limits_{i = 1}^{\nu_M}\; \left[ \frac{-\|\nu _i \|\;\pi} {l_M} \;    \right] .
\end{eqnarray}   

It can be observed that $H^{'}_{PV}$ can have any sign. However,  $H^{''}_{PV} $ is necessarily $\le 0$. Moreover,  if the EV  of type $\nu$ is evenly degenerate, it has to be recognized that such   EVs are equivalent to a $c.c.$ EV, with zero imaginary part. Thus,   $ {\frac{{\nu _i }}{{l_M}}\ln \frac{{\nu _i }}{{l_M}}}$ 
 $\equiv$                
$ - \frac{\| \nu _i ^{'}\|}{l_M}   \;ln  [\frac{\| \nu _i ^{'}\|}{{l_M}}] \pm \frac{\| \nu _i ^{'}\|}{l_M}  i\; [ \pi ]$.  Therefore, the summation over such degenerate EVs  is equal to  $-2\;  \frac{\| \nu _i ^{'}\|}{l_M} \;ln [\frac{\| \nu _i ^{'}\|}{l_M}] $ and leads to a zero imaginary part contribution.

Note also that if  the  Principal Value (PV) space  is  rather the PV' space such that   $\Phi_i$ $\in$ [-$\pi, \pi$[, one would obtain  $ {\frac{{\nu _i }}{{l_M}}\ln \frac{{\nu _i }}{{l_M}}}$ 
 $\equiv$                
$ - \frac{\| \nu _i ^{'}\|}{l_M}   \;ln  [\frac{\| \nu _i ^{'}\|}{{l_M}}]  + i\; [ \pi ]\; \frac{\| \nu _i ^{'}\|}{l_M}$, instead.  Appropriate modifications would have to occur when writing  $H^{'}_{PV'}$  and $H^{''}_{PV'}$. Yet,   the contribution of evenly degenerate   negative EVs  would still lead to a vanishing contribution to $H^{''}_{PV'}$. 


To verify the above formulae and conclusions, it can be also usefully checked that the $\nu$ and $\mu$ cases can be obtained as special cases of the $\lambda$ case.

\section{Data}\label{exempldata}

Let us consider a numerical example covering various cases among complex  (= complicated) networks.These have   been used to depict the characteristics of various abstract systems.  
  The network nodes can be motifs or agents, while the  links  can be directed or undirected, weighted or not.

Among  directed networks
  \cite{ch1,Compl1,Sinatra,CCE28.04.1789complxoptimalNetwk,araujoFFNIA,bit1,chinarailwaynetwk,AbeSuzuki,bik34,cond-mat/0501067,sfi/0412037,PRL100.08,bir2,bis2}, a special class is the citation networks \cite{ch7,PNAS107.10.13636-41-Szell-RLambi-SThur,garcia,physaGR}, which is called a digraph.  Digraphs are surveyed in \cite{BermanandShakedMonderer2008}.

  Due to the  intrinsically time dependent hierarchical  process,  the adjacency matrix representing the network is usually asymmetric, beside being a non-negative matrix.    
There is a large body of mathematical work on spectra of adjacency matrices, ranging from modern versions of Perron-Frobenius theorem for non-negative matrices \cite{Perron,Frobenius}, - up to recent results reviewed by  Brualdi
  \cite{Brualdi}  and others like  \cite{Bapat and Raghavan 1997,Berman Neumann and Stern 1989,Berman and Plemmons 1994,Minc 1988, Rothblum 2006,Senata 1981}.

For the following, let it be recalled that a  square matrix $\cal M$ is so called  irreducible if $D(\cal M)$,  the corresponding network is strongly connected,  i.e.    a path exists between any couple of nodes. Otherwise $\cal M$ is  said to be reducible. Let $\cal M$ $\ge$  0 be an irreducible $n$ x $n$ matrix. Then,  $\cal M$ has a positive real eigenvalue equal to its spectral radius $\rho(\cal M).$ 
 An irreducible nonnegative matrix $\cal M$ is said to be primitive, if the only eigenvalue of $\cal M$ of modulus $\rho(\cal M)$  is $\rho(\cal M).$ 
An irreducible nonnegative matrix $\cal M$ is said to be cyclic of index $k$   $>$ 1, if it has $k$ eigenvalues of modulus equal to $\rho(\cal M).$

To illustrate the arguments leading to the CIE theory, we have therefore selected a citation network, previously studied along other lines \cite{garcia,physaGR}, but presenting a set of adjacency matrices containing quite a variety of features, in particular with respect to the EV distributions.

 It might be useful to the reader to consider the case of the smallest asymmetric adjacency matrix, with zero on the diagonal for reasons given below,  discussed   in   Appendix C. 
   Even though it might look surprising to discuss a 3x3 matrix in a modern scientific paper, the  illustrations found in the Appendix have  been found to be the most simple ones leading to the appreciation of the EV behavior of large random matrices. Indeed such matrices correspond to complicated networks from which it is not often possible to easily observe the key features.

  \subsection{A 77x77 real asymmetric matrix } \label{2comm}
  
Consider the  citation network first studied in \cite{garcia}, with adjacency matrices given in \cite{physaGR}, according to the outcome from a  Scholar Google search process. The citations  are those of agents belonging to two quite distinct communities, composed of modern creationists (most are Intelligent Design (ID) proponents (IDP), on one hand,
and Darwin's theory of Evolution Defenders (DED), on the other hand. 
 The  network  \cite{physaGR} is composed of two subgraphs, one with 37 and the other  with 40 elements, or nodes, or agents, corresponding to the  IDP and DED community, respectively.      
 The adjacency matrices can be summarized as

\begin{equation} \label{M0}
M_0\equiv\left( \begin{tabular}{ll} $ C_0$  &   $A$   \\  $ B $ & $D_0$ 
\end{tabular} \right)\; \equiv \left( \begin{tabular}{ll} $ C_0$  &   $0$   \\  $ 0 $ & $D_0$ 
\end{tabular} \right)+  \left( \begin{tabular}{ll} 0 & $A$  \\  $B$ & $0$\end{tabular} \right).
\end{equation} in which a matrix element $m_{ij}$ takes the value 1 or 0 depending on
whether or not  a citation of $i$ by $j$  has taken place, 
as recorded and explained in ref. \cite{garcia,physaGR}. The matrices   $C_0$  (37x37) and $D_0$ (40x40)  indicate whether  agents of community $i$  have been quoted by others of the same community $i$. In contrast,  $F_0$, i.e.
\begin{equation} \label{F}
F_0=\left( \begin{tabular}{ll} 0 & $A$  \\  $B$ & $0$
\end{tabular} \right).
\end{equation}  emphasizes links $between$  different communities, i.e.    agents of community $j$  quoting those of community $i(\neq j)$; $i \leftarrow j$.
 $A$ and $B$ are obviously $rectangular$ matrices describing inter-community  links. We emphasize with the $_0$ index     that   all diagonal terms in $M_0$, $C_0$, and $D_0$
are $0$, i.e. we are $not$ considering any self-citation, i.e. $m_{ii}=0$. In brief, there are  91, 71, and 119 links, in IDP, DED and inter-community ones, respectively.  
All  adjacency  matrices, as well as $M_0$ and $F_0$  are markedly asymmetric and contain only real and positive numbers, 0 or 1; see  \cite{physaGR} for the list of all finite matrix elements.

Moreover,   since each square matrix  $M_0$, $C_0$, $D_0$, $F_0$ has non-negative elements,  the Perron-Frobenius theorem, see above,  states that there exists  a non negative eigenvalue greater or equal  {\it in absolute value} than all other eigenvalues (and  its 
  corresponding eigenvector has non-negative components) \cite{Meyer 2000,Gantmacher 2000}. 

We have   tested the hypothesis of irreducibility of these matrices.
Since the property of irreducibility  is equivalent to the property 
of the adjacency matrix corresponding to a strongly connected network, we have tested  the irreducibility through the algorithm of Tarjan \cite{Tarjan} 
It  occurs that each matrix is  reducible  (i.e.  not irreducible). In fact,  beside 
a giant strongly connected component, 
there are many other (even $single$) units that stand alone as if  strongly connected components for its neighborhood.
This is due to the fact that the "sample" reports authors that either only quote or
are only quoted. Therefore, reversal links are missing, increasing thereby  the number of strongly connected components. Therefore,   extensions of the Perron-Frobenius theorem to irreducible matrices cannot be applied since each matrix   $M_0$,   $C_0$,   $D_0$, $F_0$ reducible. 

Of course,  the Perron-Frobenius theorem for irreducible matrices could be applied  on each single strongly connected component, - locally, w mean. However, it seems that this would give a (too) local result on 
the eigenvalue for all the  short dimension sub-matrices. We have considered that it would not be useful for the present purpose, - nor to the global task
of understanding the relevance of authors, - since the property of irreducibility is
only interesting, in our opinion, for considering the strongly connected components. However, the strongly connected components may help in understanding the existence of  "hubs", clusters and clubs, - if such is a question.  
Therefore, the Perron-Frobenius theorem, applied in its version for non-negative
matrices only, indicates that there may exist eigenvalues of the same absolute value as the maximal one; moreover the maximal eigenvalue may not be a simple root  of the characteristic polynomial, can be zero (and the corresponding eigenvector does not need to be strictly positive).

 The EVs of the above 4 matrices,  $M_0$, $C_0$,  $D_0$, and $F_0$,  have been computed. The EVs of interest are given in Tables 1-3.  
 Some information "summarizing" the structure of the EVs distribution is given in Table 5.  The values of the CIE  are also given in Table 4, from which the real part, imaginary part, absolute value and phase factor of $H$ could be deduced depending on the defining Riemann  sheet.

\subsection{Discussion}\label{discussion}

First, note  some apparent similarity between the characteristics of the whole network and of the two subnetworks from  Table 4. The ratios between the number of directed (DL), undirected (UL) or total number  (L) of links are rather similar, - the DL/L ratio being a little bit larger for $C_0$.  

Moreover, each largest EV is real and positive. The second largest EV can be either positive, like in  $M_0$ and $D_0$, or complex, like in $C_0$ and $F_0$.

 Recall that the real part of the main  eigenvalue of  an  adjacency matrix is usually considered to be a measure of centrality of some node, like a leading agent. What is new here concerns  the imaginary part  which, as we consider, indicates  the relative  (inverse) scale or range of influence of such a leader.

The information entropy, deduced from the  calculated EVs,  enhances some difference between the networks, going beyond consideration on the "main node", or "state" in a thermodynamic language.
The real and imaginary parts of    information entropy  values for   the  matrices are presented in  Table 4.  
First, one notes that $H_1$  values  are all of the same order of magnitude. Next, it is  apparent that  $\| H\| $ are quasi similar  whatever the used Riemann sheet, but more interestingly that the $M_0$ and $D_0$  values differ  from those of $C_0$ and $F_0$. Yet,  the $D_0$ $\| H\| $ value is very small and quasi  equal to 0,  in using the PV space.

 $H^{'}$   is markedly different  for $C_0$  and for the other matrices. In fact, the real part of the IDP network $entropy$ $H$ is the only one to be positive, - but small.  The other real parts are  negative, in both spaces.  
 In all cases,  $H^{''}$ is negative.   
 
 Recall that the size of the matrices has   been taken into account through the $l_M$ factor in the EV normalization. 
 
Rather than spending more time on the numerical values, it  seems more useful to stress the origin of such a complex information entropy as arising  through the network  structure. The key feature appears to be the ensemble of  so called simple transitivity paths, as illustrated in App. C on small networks.   This observation might also help in future deep mathematical work in  order to sort out the property of EV distributions of random matrices. {\it In fine}, the complex part of the entropy truly emphasizes the delay in  information, the citation sequence, and range of  information, the connected nodes.
 In so doing, it seems that  the CIE has brought   a   vision of  directed networks  different from other measures and other previous discussions, like one in \cite{PRE80.09.046110rdmgraph}, but considering acyclic networks.

 \section{Conclusions}\label{concl}

 Directed networks are very common.  Citation networks belong to a huge subclass of those.  It is   common knowledge also that opinion formation demands information exchanges which are thereby necessarily "directed" between agents.
When representing such networks through adjacency matrices, it is apparent that such matrices are necessarily asymmetric.  
 
  This paper, on one hand, introduces a  technique  in order to obtain some insight into directed networks.  It appears that one can consider a network information entropy through the link distribution on which information is exchanged. Through an analogy with Boltzmann entropy in usual statistical mechanics, one can observe that in order to get more insight  on the network entropy, one can calculate the whole set of eigenvalues of the adjacency matrix of the network.  This is similar to consider the set of discrete values of a transfer matrix in quantum or statistical mechanics. However, the resulting information entropy turns out to be a complex mathematical feature. It needs some interpretation\footnote{A reviewer suggested that "it would be fair, to indicate which branch is ÒbetterÓ or ÒworseÓ in possible applications/interpretations". This interesting point, however, sends back the reader to wonder what Riemann sheet is used in numerical algorithms, - ... different ones, as we have alas observed reconciling various calculations, whence  inducing the  extensive reports in Sect. 2.2.1 and 2.2.2, surely serving as warnings}. The latter can be based on the free energy concept.

Starting form the notion of equilibrium free energy,   Zweger   \cite{TCFE} attempted a dynamical interpretation of  a classical complex free energy, in 1985.
He pointed out that the problem is to determine a characteristic "relaxation" time  for some process in which the dynamical (Langevin or Fokker-Planck) equation is connected to some Hamiltonian or some corresponding  transfer matrix. A probability current can be written, in fact, in terms of some unstable mode times an equilibrium factor which is the imaginary part of the free energy \cite{Langer1969}. Usually \cite{TCFE,Cook}, the imaginary part of the free energy (or largest eigenvalues) give some information about  the "nucleation stage" of the dynamics.   The real part, of course, determines the equilibrium energy state.    
 
 We propose that another, though related approach, can be considered.  Instead of some "relaxation time", one may consider  the "spatial aspect of the phenomenon", e.g. through some correlation range  length $\xi$. This approach makes some sense,in particular for networks, analytically described through some Hamiltonian or transfer matrix. Moreover, such a spatial scale   introduction may remind of some analogy with Discrete Scale Invariance (DSI)  (or lack of DSI) feature \cite{DSI,DSphysrep}. In fact, this DSI leads to complex dimensions and complex critical exponents.

  The illustrating example  implies two distinct communities, with  markedly different opinions : the  Neocreationist and Intelligent Design  Proponents (IDP)  on one hand, and the Darwinian Evolution Defenders (DED) on the other hand.   These are communities for which an opinion consensus  can be hardly expected.  
  It appears that for the whole set of agents, two agents  ("states") are markedly dominating. They seem to belong to the DED community. In contrast, the IDP community has only one main "state". Interestingly, the same is true for the inter-community " information exchange phenomenon" for which there is only one main dominating state.

In summary,  we have presented an original work extension of  $IE$, connecting the CIE method,   to sound statistical mechanics. By examining, different eigenvalues of asymmetric matrices, - sometimes   complex eigenvalues, yet starting with the largest ones, one can describe an $IE$, - like if in thermodynamics, one describes a free energy in terms of eigenvalues of some Hamiltonian. Thus, one not only obtains the "basic" free energy, but also corrections due to some underlying scale  structure. Moreover  considerations on the   mathematical   form of  the $IE$, i.e. its real and imaginary part, when they exist, allow to emphasize characteristics, which we attribute to the leadership range. This has been exemplified by considering a network with two specific communities having different opinions, exchanged through citations.
  
   \bigskip
 
{\bf Note added at the completion of this report}:  Another argument in favor of studying asymmetric matrices can be mentioned. Indeed, during  the process of finishing up  the present work, for 
submission,  a paper, submitted to to EPJB,  occurred on arXives \cite{Lucaasymmcorr} entitled {\it  Asymmetric correlation matrices: an analysis of financial data}. Due to the asymmetry in time delayed correlations between financial time series, it is indeed also of interest to extend the spectral analysis to the realm of complex eigenvalues, - as first attempted in  \cite{GinOE}.  Though different in essence, such works and the present one indicate that one should not be stacked to studying only systems within real algebra.

 \vskip 0.5truecm
 {\bf Acknowledgements}

GR thanks the COST Action MP0801   for the STSM 4475 grant, 	allowing her stay at the University of Liege in  Feb. 09. MA thanks the COST Action MP0801   for the STSM 6698 grant, 	allowing his stay at the University of Viterbo, in    Sept. 2010.

   \newpage
{\bf Appendix A.} {\bf Normalization considerations}\label{Normalization}  \vskip 0.5truecm

In the main text,  the  Theil index form \cite{Theil65}
\begin{equation} \label{Theilform}
 \frac{{ \lambda _i } }{{< \lambda _i >}}    \sum\limits_{i = 1}^{l_M}\;  \frac{{ \lambda _i } }{{< \lambda _i >}} , \end{equation}
has been adapted from its original writing, with real numbers, to one involving complex numbers, but the more so better appropriate for the $ IE$,  i.e., 
\begin{equation} \label{Protoform}
 \frac{{ \lambda _i } }{{l_M}}   \sum\limits_{i = 1}^{l_M}\;  \frac{{ \lambda _i } }{{l_M}}, \end{equation} with the "normalization"  $l_M$.   Indeed, it would have been inappropriate to use the Theil index original normalization, since  $< \lambda _i > =0$, because  $  \sum\limits_{i = 1}^{l_M}\;    \lambda _i  =0 $.

 \vskip 0.5truecm
  
{\bf Appendix B.} {\bf Tsallis entropy} \label{Tsallisentropy}  \vskip 0.5truecm

Tsallis  \cite{tsallis1,tsallis2,tsallis3}  proposed that a large category of  systems may be treated by a similar
formalism,  but where a more general entropy measure is defined by

\begin{equation} \label{tsallis} S_q =k_B \frac{(1-\sum_{i} p_i^q)}{q-1}
\end{equation}
which depends on the real parameter $q$ and which reduces to the Shannon
entropy for   $q\rightarrow$1.  Along the lines in the main text, it is tempting to  define a $_qIE$ as
\begin{equation} \label{tsallisq} 
H_q \equiv\; 1+\;\frac{k_B}{q-1}  (1-\sum_{i} p_i^q) \equiv\;1+ \frac{k_B}{q-1}  \left(1-\sum_{i} [\frac{ \lambda_i}{l_M}]^q \right)
\end{equation} thereby accepting a complex-valued $IE$ in Tsallis sense.

Tsallis theory is sometimes referred  to in the literature as no-extensive statistical mechanics, in contrast with the
extensivity of the Shannon entropy. For general $q$, a proper extremisation of Eq.(\ref{tsallis}) leads to generalized canonical distributions, often called Tsallis distributions,
\begin{equation} \label{thermo} f_q(x)=\frac{e_q^{-\beta^{'} x}}{Z_s(q)}
\end{equation}
where $x$ denotes the energy of the system,  
  $ Z_s(q) \equiv  \Sigma_{\nu=1}^{N_s}  e^{-  {\cal H}(s, \nu)^q}$, and $e_q^x$ is the
$q$-exponential function defined by
\begin{equation} e_q^x \equiv (1+(1-q) x)^{\frac{1}{1-q}}.
\end{equation}

 One should also note that  Tsallis  formalism draws a direct
parallelism with the equilibrium theory, where $\beta^{'}$ plays the
role of the inverse of a temperature, and $Z$ that of a partition
function.  However Tsallis  and others  \cite{tsallis1,tsallis2,tsallis3} have often insisted ion the connexion between $q$- and non-equilibrium effects. One might consider  connecting  the above $IE$ to  Tsallis considerations, in further work; see already \cite{APhPB35.04.871}. Recall that a $q-$Theil index has been already introduced \cite{JM40,IdA}.


\vskip 0.5truecm

     {\bf Appendix C.}\label{3x3matrix}   {\bf A 3x3 matrix}   \vskip 0.5truecm
 
 Let a state Hamiltonian  be described by a 3x3 matrix
   \begin{equation} \label{3x3H}
\cal H =
\left( \begin{tabular}{llll} $ H_{11}$  &   $H_{12}$  &   $H_{13}$   \\ 
 $H_{21}$  &$H_{22}$  &   $H_{23}$ \\
 $ H_{13}$  &   $H_{23}$  &   $H_{33}$  
\end{tabular} \right)\;.
\end{equation}
To remain within a "no self-citation" scheme, let all the diagonal elements be equal to zero, i.e. 
$ H_{11} =H_{22}=   H_{33} $ = 0, and call this "new" matrix ${\cal H}_0$. Moreover, let all non diagonal elements be equal  to either 1 or 0. These "reductions" are made in the spirit of tying the present subsection to the main text, involving a  ("large") citation network for which the (adjacency) matrix has elements taking only  a 1 or 0 value.

Recall that the EVs of any 3x3 matrix are solutions of the cubic equation
   \begin{equation}    \label{cubic}
   - \lambda^3 +  \lambda^2   \;    tr ({\cal H}_0)  
  + \frac{\lambda}{2}
   [tr ({\cal H}_0)^2- tr ^2({\cal H}_0)] + det ({\cal H}_0)=0. 
 \end{equation}  
 
By "construction", $    tr (\cal H$$_0)= tr ^2(\cal H$$_0) =0. $ Moreover, one easily obtains
 that $tr (\cal H$$_0)^2/2=  H_{12}H_{21}
+ H_{13}H_{31} + H_{23}H_{32} $, and  $det (\cal H$$_0) =H_{13}H_{32}H_{21}+H_{31}H_{12}H_{23}$. Therefore, only 7 types of cubic equations, as Eq.(\ref{cubic}),  have to be considered

\begin{itemize}
\item type I \;: $-\lambda^3 +3\lambda+2=0$
\item type II \;: $-\lambda^3 +2\lambda+1=0$
\item type III\;: $-\lambda^3 + \lambda+1=0$
\item type IV\;: $-\lambda^3 + \lambda+0=0$
\item type V \;: $-\lambda^3 +2\lambda+0=0$
\item type VI\;: $-\lambda^3 +0\lambda+1=0$
\item type VII\;: $-\lambda^3 +0\lambda+0=0$
\end{itemize}

It is somewhat easily deduced that only type III and type VI lead to complex eigenvalues. 
 Both types have one positive real root.
  Note that the type I cubic equation has a real negative (of course evenly) degenerate root = -1, {\it requesting special attention} when calculating $H$; see end of Sect. \ref{CIEM}. 

\begin{figure}
 \caption{Networks of  triplets, according to Pajek Manual notations,  corresponding to different types of cubic equations as given in the main text; only one network is shown per type of equation; the others can be obtained by permutation.} 
\includegraphics[height=24cm]{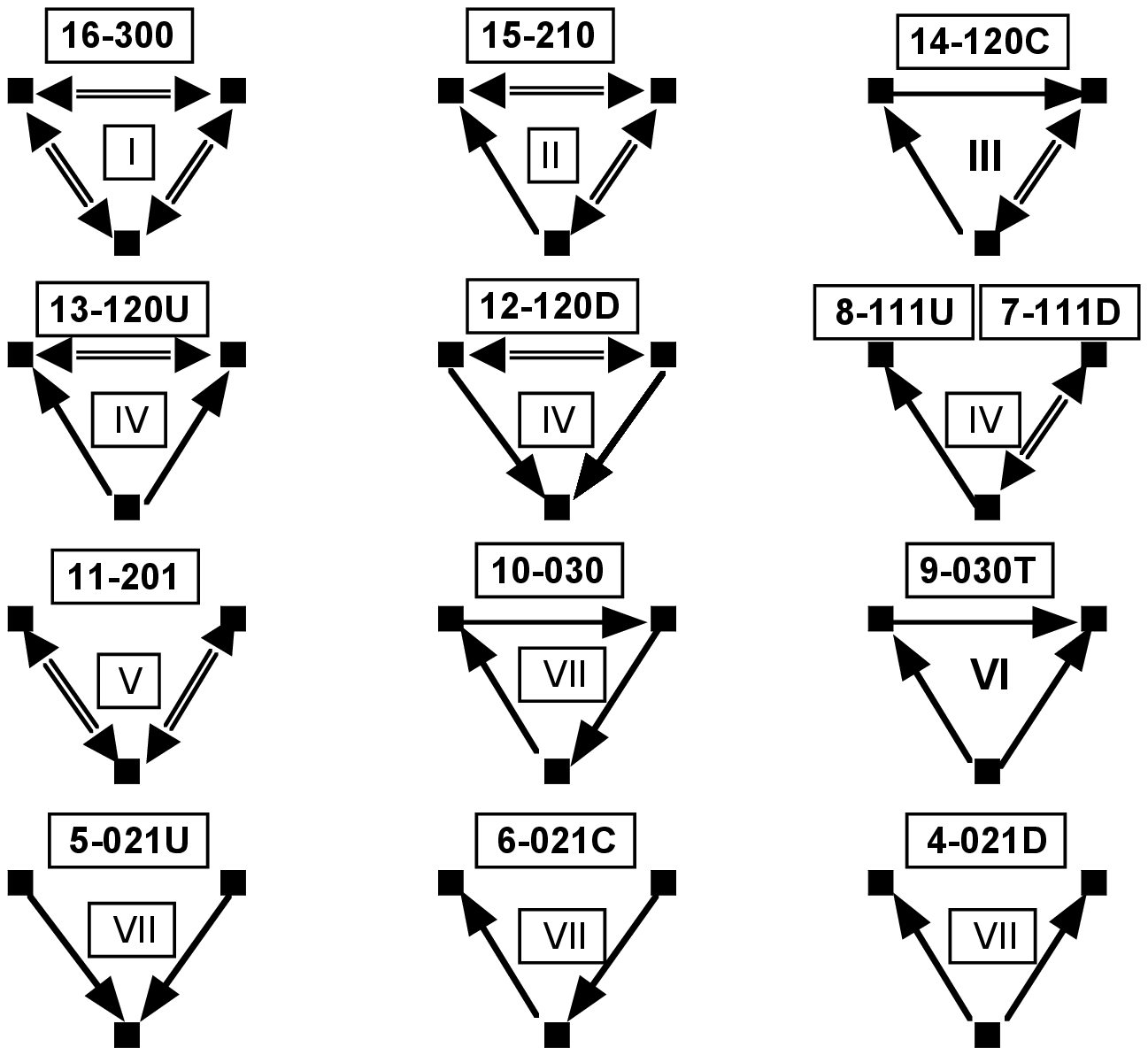}  
\label{fig:triads}
\end{figure}

The networks made of three nodes corresponding to such cubic equations are illustrated in Fig. \ref{fig:triads}, - one network is displayed for each case only; the others are easily and readily deduced by permutation of bonds \cite{MOO,MSIKCA,MIKLSA}. On one hand, this illustrates well why type I  has degenerate eigenvalues. On the other hand, the  complex eigenvalues (type III and type VI) are now understood as arising from the transitivity relationship, corresponding to 14-120C and 9-030T triads, in Pajek Manual notations \cite{pajek}. 

Even though it might look surprising to describe 3x3 matrices in a modern scientific paper, the present illustration has been found necessary as the most simple one leading to some appreciation of the EV behavior of larger  random matrices.

\newpage
 \begin{figure}\caption{Distribution of the eigenvalues  (EVs) of the matrix $M_0$ in the complex plane }
\includegraphics[height=20cm]
{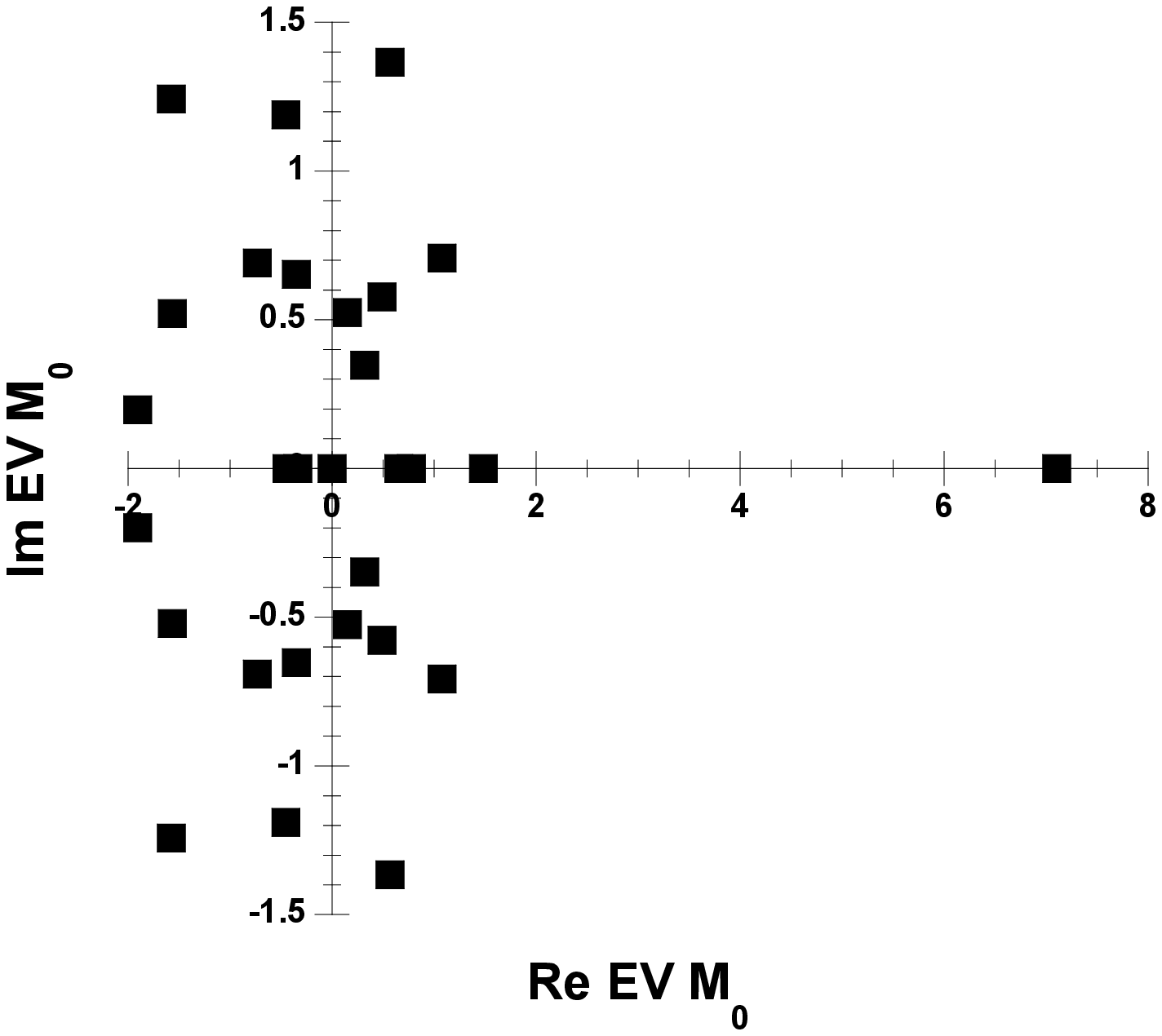}
 \label{fig:EVM0}
\end{figure}

 \begin{figure}\caption{Distribution of the eigenvalues  (EVs) of the matrix $C_0$ in the complex plane } 
\includegraphics[height=20cm]{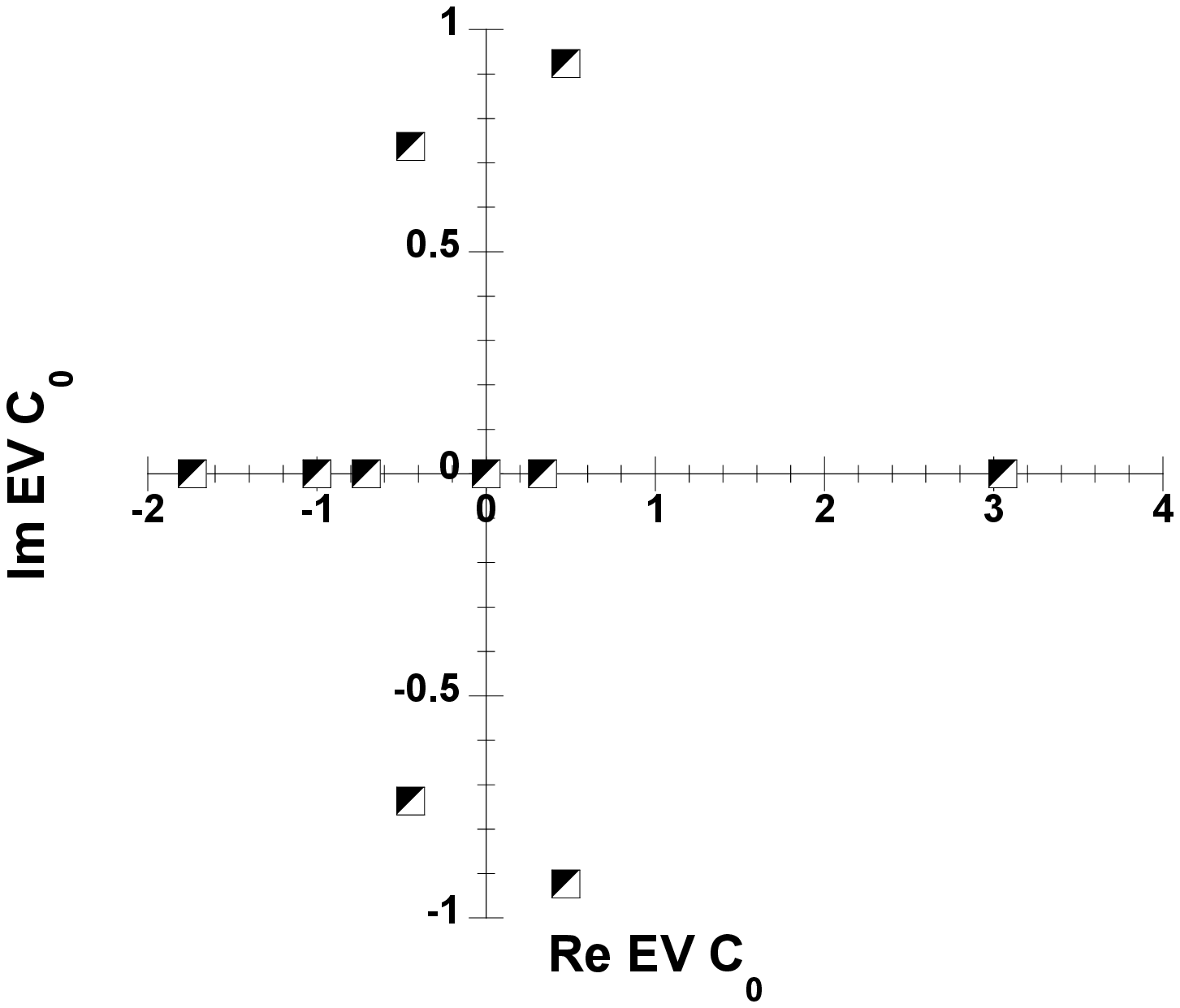}
\label{fig:EVC0}
\end{figure}

 \begin{figure}\caption{Distribution of the eigenvalues  (EVs) of the matrix $D_0$ in the complex plane }
\includegraphics[height=20cm]{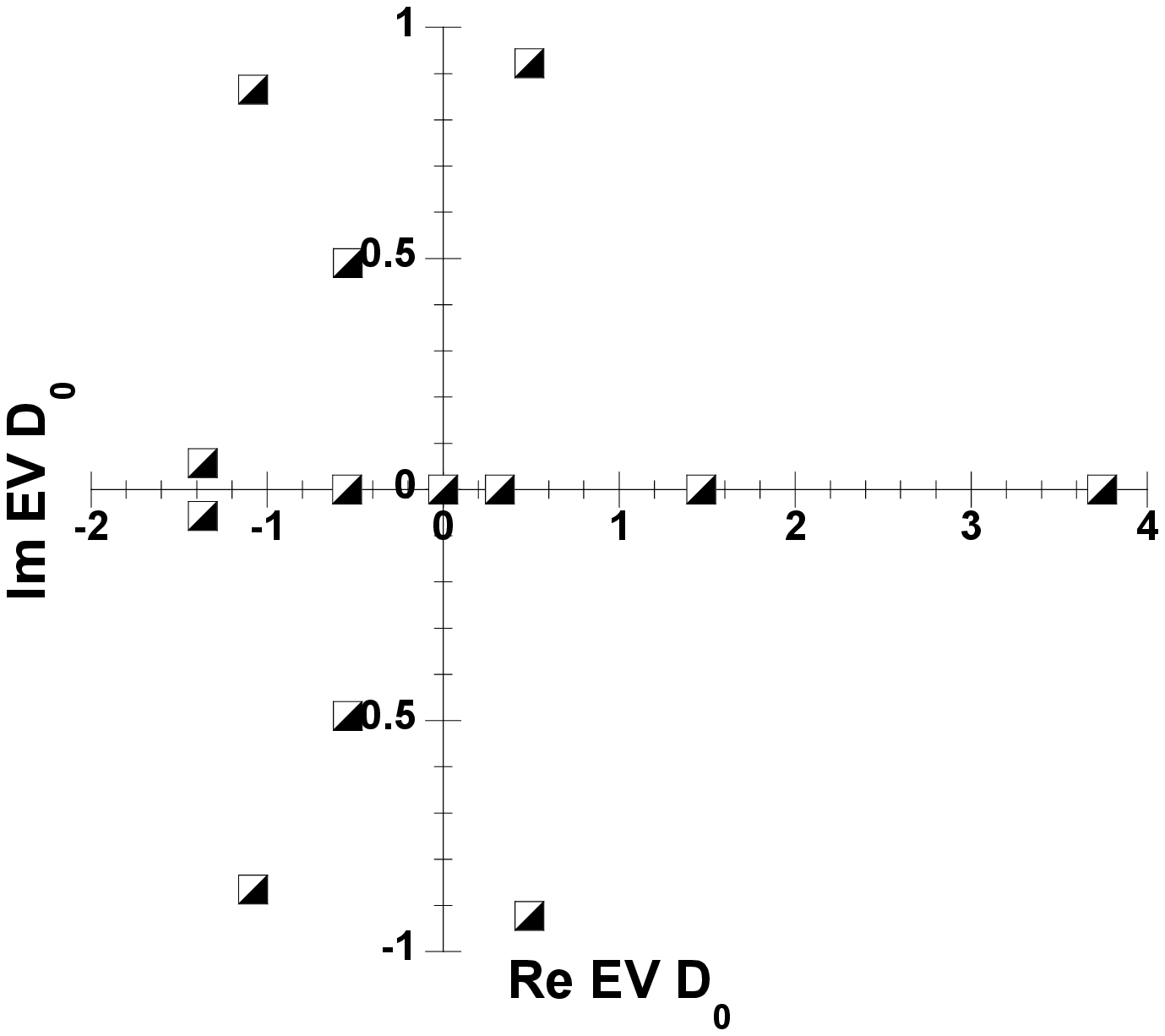}
 \label{fig:EVD0}
\end{figure}

 \begin{figure} \caption{Distribution of the eigenvalues  (EVs) of the matrix $F_0$ in the complex plane }
\includegraphics[height=26cm]{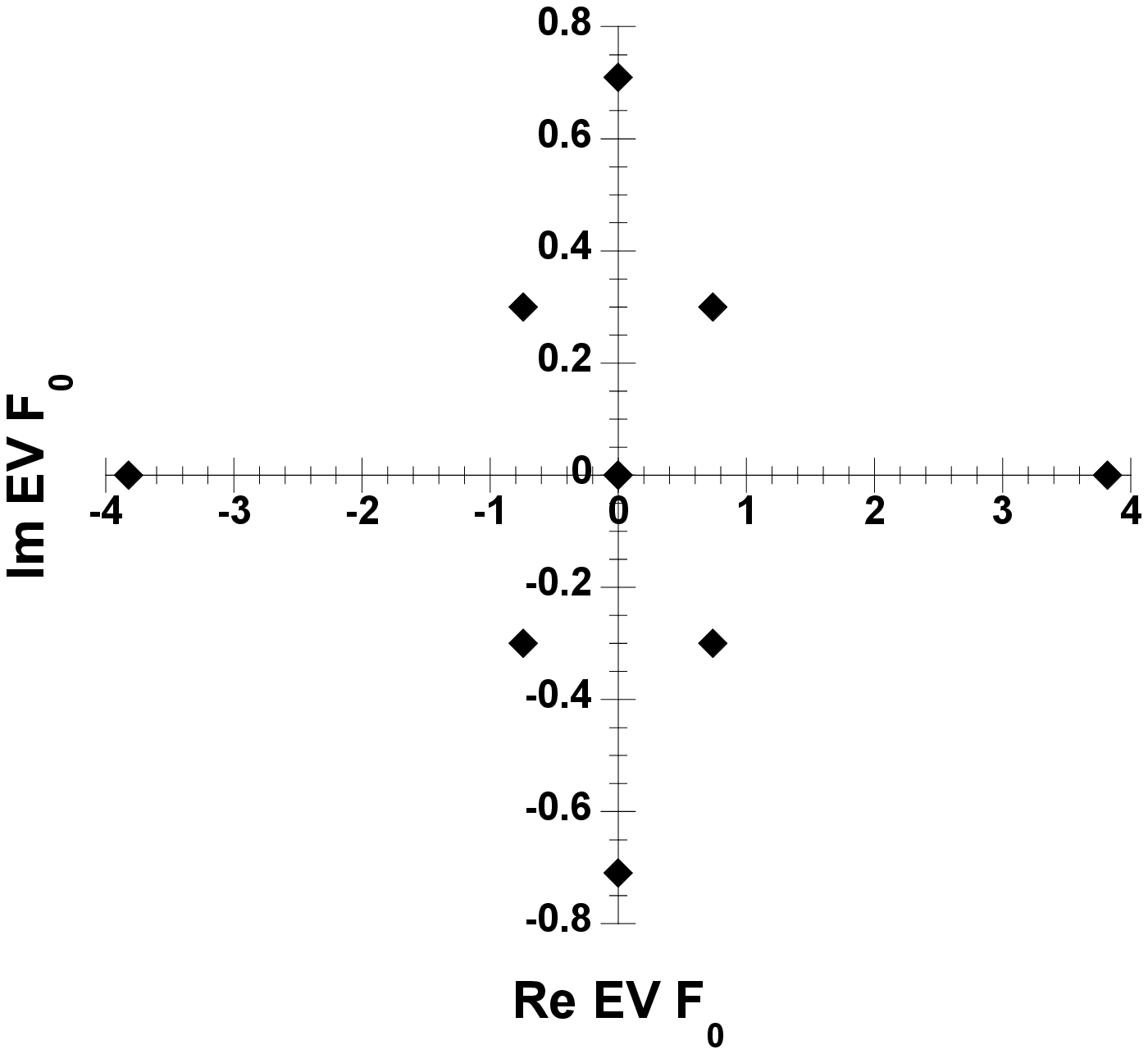}
 \label{fig:EVF0}
\end{figure}


     \begin{table}[t]
           \begin{center} 
           \begin{tabular}{|c|c|c| |c|c|c|  }
\hline     $M_0$&$.$ &$.$ &$M_0$&$.$ &$.$   \\
\hline\hline  $ EV\;rank$  &$ Re $ &$Im$   &$ EV\;rank$  &$ Re $ &$Im$   \\ 
\hline\hline $1$&$ 7.1039	$&-&$62 $ &$ -0.0000	$&$-0.0000  $\\   
\hline $2$&$ 1.4849$  &$	-$& $63$&$-0.3350$&$-$ \\   
\hline $3$&$ 1.0758$  &$	0.7077$&$64$ &$-0.3461$&$0.6531$\\    
\hline $4$&$ 1.0758$  &$	-0.7077$& $65 $ &$-0.3461$&$-0.6531$\\    
\hline $5$&$ 0.7763$  &$	-$&$66 $ &$-0.4422$&$-$\\      
\hline $6$&$ 0.5693$  &$	1.3660$&$67 $ &$-0.4524$&$1.1900$\\
\hline $7$&$ 0.5693$  &$	-1.3660$&$68 $ &$-0.4524$&$-1.1900$\\    
\hline $8$&$ 0.4919$  &$	0.5775$&$69$&$-0.6560$&$-$ \\   
\hline $9$&$ 0.4919$  &$	-0.5775$&$70$&$-0.7323$&$0.6917$ \\   
\hline $10$&$ 0.3214$  &$	0.3474$&$71$&$-0.7323$&$-0.6917$ \\     
\hline $11$&$ 0.3214$  &$	-0.3474$&$72$&$-1.5661$&$0.5212$ \\   
\hline $12$&$ 0.1506$  &$	0.5245$&$73$&$-1.5661$&$-0.5212$ \\     
\hline $13$&$ 0.1506$  &$	-0.5245$&$74$&$-1.5739$&$1.2422$ \\ 
\hline $14$&$ 0.0000$  &$	0.0000 $&$75$&$-1.5739$&$-1.2422$ \\      
\hline $... $ &$ 0.0000	$&$0.0000  $&$76$&$-1.9041$&$0.1983$ \\  
\hline $61$ &$ -0.0000	$&$-0.0000  $&$77$&$-1.9041$&$-0.1983$ \\    

\hline
 \end{tabular}
    \label{EVM0 } \end{center}\caption{Real ($Re$) and Imaginary ($Im$) part of each eigenvalue (EV)  of the $M_0$ matrix ranked in increasing rank order according to the  $Re$ part.  Not all EVs $\equiv 0$ are  given.}  \end{table}

     \begin{table}[t]
           \begin{center} 
           \begin{tabular}{|c|c|c||c|c|c|  }
\hline     $C_0$&$.$ &$.$ & $D_0$&$.$ &$.$   \\
\hline\hline  $ EV\;rank$  &$ Re $ &$Im$& $ EV\;rank$  &$ Re $ &$Im$   \\
\hline\hline $1$&$  3.054$&$-$&$1$&$ 3.744$&$-$  \\   
\hline $2 $ &$ 0.4714$&$	0.9238 $&$2$&$1.4677	$&-\\   
\hline $3 $ &$ 0.4714$&$	-0.9238  $&$3$&$0.4907	$&0.9230\\   
\hline $4 $&$ 0.3349$&$	-  $&$4$&$0.4907	$&-0.9230\\   
\hline $5 $ &$ 0.0000	$&$0.0000  $&$5$&$0.3213	$&-\\ 
\hline $6 $ &$ 0.0000	$&$0.0000  $&$6$&$0.0000	$&0.0000 \\   
\hline $... $ &$ 0.0000	$&$0.0000  $&$...  $ &$ 0.0000	$&$0.0000  $\\   
\hline $32 $ &$ 0.0000	$&$0.0000  $&$32 $ &$ 0.0000	$&$0.0000$\\   
\hline $33$ &$-0.4441$&$ 0.7369$&$33$&$0.0000	$&0.0000\\   
\hline $34 $ &$-0.4441$&$	-0.7369$&$34$&$-0.5423$&	0.4902\\   
\hline $35 $ &$-0.7072$&$-$&$35$&$-0.5423$&	-0.4902\\   
\hline $36 $ &$-1.0000$&$	-$&$36$&$-0.5441$&	-\\  
\hline $37$&$-1.7363$&$-$&$37$&$-1.0789	$&0.8657\\  
\hline &&&$38$&$-1.0789	$&-0.8657\\  
\hline &&&$39$&$-1.3640$&	0.0572\\ 
\hline &&&$40$&$-1.3640	$&-0.0572\\
\hline
 \end{tabular}
    \label{EVC0D0} \end{center}\caption{Real ($Re$) and Imaginary ($Im$) part of each eigenvalue (EV)  of the $C_0$   and $D_0$ matrix ranked in increasing rank order according to the  $Re$ part.  Not all EVs $\equiv 0$ are  given.}   \end{table}
         \begin{table}[t]
           \begin{center} 
           \begin{tabular}{|c|c|c| |c|c|c|  }
\hline     $F_0$&$.$ &$.$ &$F_0$&$.$ &$.$   \\
\hline\hline  $ EV\;rank$  &$ Re $ &$Im$   &$ EV\;rank$  &$ Re $ &$Im$   \\ 
\hline\hline $1$&$ 3.8191	$&-&$73 $ &$- $&$0.7097$\\    
\hline $2$&$ 0.7411$  &$	0.2997$& $74$&$- $&$-0.7097$ \\    
\hline $3$&$ 0.7411$  &$	-0.2997$&$75$&$-0.7411$&$0.2997$ \\   
\hline $4$&$ 0.000$  &$0.000$&$76$&$-0.7411$&$-0.2997$ \\   
\hline $...$&$ 0.000$  &$0.000$& $77$&$-3.8191$&$-$ \\     

\hline
 \end{tabular}
    \label{EVF0 } \end{center}\caption{Real ($Re$) and Imaginary ($Im$) part of each eigenvalue (EV)  of the $F_0$ matrix ranked in increasing rank order according to the  $Re$ part.  Not all EVs $\equiv 0$ are  given  
    }  \end{table}

 \begin{table}
           \begin{center}
           \begin{tabular}{|c|c|c|c|c|c|c|  }
\hline   $number$ $of$  &&$M_0$ &$C_0$ &$D_0$ &$F_0$    \\
 \hline\hline  $nodes$ ($l_M$)&&$  77$&$37$  &40&$77$    \\   
\hline $links $ &&$ 281$&$91$  &71&$119$ \\    
\hline $directed $ $links$&  &$ 219$&$79$  &51&$89$   \\   
\hline $undirected$ $ links$&&$31$&$6$  &10&$15$   \\  \hline
\hline $finite$ $EVs$&&$28$&$9$  &12&$8 $    \\   

\hline $EV\equiv1$&&$0$&$0$  &0&$0$  \\   
\hline $EV\equiv0$&&$49$&$28$  &28&$69$  \\   
\hline $EV\equiv -1$&&$0$&$0$  &1&$0$  \\  
\hline  $real, deg.$ $EV\neq0$&& 0 &0  &0&0   \\   
\hline $\rho$ $ $  $Re>0$&&$3$&$2$  &3&$1$   \\   
 \hline $\nu$ $  $  $Re<0$&&$3$&$3$  &1&$1$   \\   
\hline $\mu$ $ $ $ Im >0$&&$0$&$0$   &0&$1$    \\   
\hline $\mu$ $ $ $ Im <0$&&$ 0$&$0$  &0& 1    \\   
\hline $\lambda$, $c.c.$ $EV$&& 11x2 &2x2  &4x2&3x2   \\     \hline
 \hline    ($\lambda_{1} $/$l_M$) ln[$\lambda_{1} $/$l_M$]  && $ -0.2199$ &$-0.2059$  &-0.2217& $-0.1490$  \\   
\hline$H_1$&&$0.7801$&$0.7941$  &0.7783&$0.8510$  \\   
\hline\hline  $H^{'}_{TC}$  &&$+0.2655$&+1.0731 &$+0.3366$  &   $ +0.8671$   \\   
\hline $H^{''}_{TC}$  &&$ -0.4568$&$-0.2877$  &$-0.4346$&$-0.2132$   \\   
\hline $H^{'}_{PV}$ &&$-0.3889$&+0.7911&$-0.03040$    &+0.7603\\   
\hline $H^{''}_{PV}$ &&$ -0.1328$&$-0.2924$  &$-0.04273$&$-0.2422$  \\   
\hline

 \end{tabular} \label{M0C0D0F0}
    \end{center} \caption{Pertinent characteristics of  studied matrices 
     } \end{table}

\end{document}